\documentclass{PoS}

\newcommand{\be}{\begin{equation}} 
\newcommand{\en}{\end{equation}}
\newcommand{\bea}{\begin{eqnarray}}
\newcommand{\ena}{\end{eqnarray}}

\newcommand{\hbo}{\hbox to 1 true cm {\hfill } } 
\newcommand{\tr}{\hbox{tr}}

\title{Towards a density of states approach for dense matter systems}

\ShortTitle{Towards a density of states approach for dense matter systems}

\author{\speaker{Kurt Langfeld}%
         \thanks{This work is supported by STFC under the DiRAC framework. 
We are grateful for the support from the HPCC Plymouth.}\\ 
Research Centre for Mathematical Sciences, 
        Plymouth University, Plymouth PL4 9DE, UK, 
        E-mail: \email{kurt.langfeld@plymouth.ac.uk}} 
\author{ Jan Pawlowski \\ Institute for Theoretical Physics, 
University of Heidelberg, 69120 Heidelberg, Germany}
\author{ Biagio Lucini \\ Department of Physics, 
Swansea University, Swansea SA2 8PP, UK }
\author{ Antonio Rago \\Research Centre for Mathematical Sciences, 
Plymouth University,  Plymouth PL4 9DE, UK }
\author{ Roberto Pellegrini \\ Dipartimento di Fisica dell'Universit\`a 
di Torino, 10125 Torino, Italy  } 

\abstract{
The density-of-states method (Phys.Rev.Lett. 109 (2012) 111601) features an
exponential error suppression and is not restricted to theories with positive
probabilistic weight. It is applied to the SU(2) gauge theory at finite
densities of heavy quarks. The key ingredient here is the Polyakov line
probability distribution, which is obtained of over 80 orders of magnitude. We
briefly address whether the exponential error suppression could be sufficient
to simulate theories with a strong sign problem.
}

\FullConference{31st International Symposium on Lattice Field Theory LATTICE 2013\\
                 July 29 - August 3, 2013\\
                 Mainz, Germany}

\begin{document}

\section{Introduction: }
First principles simulations of QCD at finite baryon densities are an
outstanding problem in particle physics due to the notorious sign problem. 
Insights into the QCD phase diagram (as a function of the chemical potential
and the temperature) might be gained by considering the heavy quark limit: 
for $SU(N_c>2)$, those theories are still hampered by a sign problem, but 
there are indications~\cite{Mercado:2011ua} that the sign problem is less
severe and solvable with recent techniques such as worm type 
algorithms~\cite{Prokof'ev:2001zz,Mercado:2012ue}, complex Langevin
techniques~\cite{Aarts:2011zn,Aarts:2012ft} or the fermion bag
approach~\cite{Chandrasekharan:2010iy}. Monte-Carlo simulations with respect
to the ``density of states''~\cite{Wang:2001ab} rather than the (potentially
complex) Gibbs factor might also be an interesting alternative to action based
simulations~\cite{Langfeld:2012ah,Bazavov:2012ex}. 

\medskip
The starting point is the continuum formulation of $SU(N_c)$ Yang-Mills theories
with the chemical potential $\mu $ for the heavy quarks with mass $m$. 
Using the heat-kernel approach, a systematic expansion of the quark
determinant in powers of $1/m$ yields the effective gluonic action (formulated
in terms of links $U_\mu$)~\cite{Langfeld:1999ig}: 
\be 
S[U] \; = \; 
S_\mathrm{YM}[U] \; + \; f \; p[U] , \hbo p[u] := \sum _{\vec{x}} P(\vec{x}) 
\label{eq:1}
\en 
where $P(\vec{x})$, the (traced) Polyakov line, and $f$ is given by: 
\be 
P(\vec{x}) = \frac{1}{N_c} \tr \; \prod _{t=1}^{N_t} U_4 (\vec{x},t), \hbo 
f = \sqrt{2} \pi^{-3/2} (mT)^{3/2} a^3 \, 
\exp \{ (\mu -m) /T \} \; , 
\label{eq:2}
\en 
and where $N_t$ is the number of links in temporal direction, $a$ is the
lattice spacing, $T = 1/N_t a$ the temperature and $S_\mathrm{YM}[U]$ is the
(Wilson) action of pure Yang-Mills theory. Note that the only dependence on
the spatial links is in the Wilson action of pure Yang-Mills theory. 
If we integrate over these links in leading order strong coupling expansion, 
we obtain an effective theory that only depends on the Polyakov lines and that 
features a nearest neighbour Polyakov line interaction. The effective theory is
the so-called $SU(N_c)$ spin model~\cite{Karsch:1985cb,Mercado:2012ue}.
We here refrain from the strong coupling expansion, but will consider 
the theory with action (\ref{eq:1}), which can be considered the {\it weak
  coupling} version of the $SU(N_c)$ spin model. For $N_c=3$, this theory 
describes full QCD with heavy quarks at finite densities. 

\medskip
Given the close relation between the Polyakov line and colour confinement, a
quantity of particular interest is the the Coleman effective potential for the
Polyakov line. This potential is defined as usual by means of the Legendre
transformation of the generating functional: 
\bea 
V(q) &=& \frac{T}{V_3} ( \mu \, q \; + \; j \, q \; - \;  \ln Z[J] \, ) , 
\hbo 
q \; = \; \frac{ d \, \ln Z[j] }{ dj} \; = \; \langle p[U] \rangle . 
\label{eq:3} \\ 
Z[j]  &=& \int {\cal D} U_\mu \; \exp \Bigl\{ S_\mathrm{YM}[U] + j 
\sum _{\vec{x}} P(\vec{x}) \Bigr\} \; . 
\label{eq:4}
\ena 
The generating functional $Z[j]$ would be obtained quite easily from the 
probability density $\rho (q)$ for finding a particular value for the 
(integrated) Polyakov line $q$: 
\be 
\rho (q) \; = \; \int {\cal D} U_\mu \; \exp \{ S_\mathrm{YM}[U] \} \; 
\delta \Bigl( q \, - \, \sum _{\vec{x}} P(\vec{x}) \Bigr) \; . 
\label{eq:5}
\en 
We find at least formally: 
$$ 
Z[j] \; = \; \int dq \; \rho (q) \; \exp \{ j \, q \} \; . 
$$
In practice, it is challenging to obtain a statistically viable result 
for the effective potential $V(q)$ (\ref{eq:3}). There are several reasons for
that: (i) Obtaining $\rho (q)$ by a standard histogram method is extremely
costly since the standard deviation of the distribution decreases with the
volume. For any value $q$ significantly different from zero, a large number 
of independent configurations is necessary, and one such configuration only
produces one entry in the histogram. (ii) The $j q $ term in the Coleman
potential (\ref{eq:3}) cancels to a great deal the $j 
\sum _{\vec{x}} P(\vec{x})$ term in the generating functional (\ref{eq:4}) 
as can be e.g.~seen from a classical approximation. Those terms are extensive 
quantities and are potentially large. This  leaves us with a poor
signal-to-noise  ratio. (iii) For $SU(N_c)>2$, $P(\vec{x})$ is complex, and
standard  Monte-Carlo method are no longer viable because of a sign
problem. In the following, we will confine us to the $SU(2)$ case for which
the sign problem is absent. We will resolve the issues (i) and (ii) by
adopting a  density-of-states approach that features an exponential error
suppression.

\section{ The density-of-states approach (LLR method) }

Let us consider the partition function $Z$ of a theory of the variable 
$\phi $ and action $S[\phi ]$: 
\be 
Z \; = \; \int {\cal D}\phi \; \exp \{ \beta \, S[\phi ] \} \; , 
\label{eq:10}
\en 
where is a solid-state physics context $\beta $ is the inverse temperature
while in quantum field theory $\beta $ is interpreted as the inverse coupling
strength. This partition function can be trivially rewritten using the 
so-called density-of-states $\rho $: 
\bea 
\rho (E) &=& \int {\cal D}\phi \; \delta \Bigl( E \, - \, S[\phi ] \Bigr) \; , 
\label{eq:11} \\ 
Z &=& \int dE \; \rho(E) \; \mathrm{e}^{\beta \, E } \; . 
\label{eq:12} 
\ena 
Knowledge of $\rho (E)$ reduces the calculation of the partition function 
to an ordinary integral. The partition function is at the heart of many
interesting quantities such as the thermal energy density, pressure, latent
heat (for theories with a 1st order phase transition) or interface tensions. 
However, the integral (\ref{eq:12}) is a real challenge: $\beta $ is usually
of order $1$ while the action $E$ is proportional to the lattice volume, which
easily reaches a $100,000$ for modest lattice sizes. It is only the lack of
states at high actions that renders the integral (\ref{eq:12}) finite. 
How do we calculate then $\rho (E)$ with a sufficient precision to obtain
significant values for $Z$? 

\medskip
\begin{figure}
\includegraphics[height=6cm]{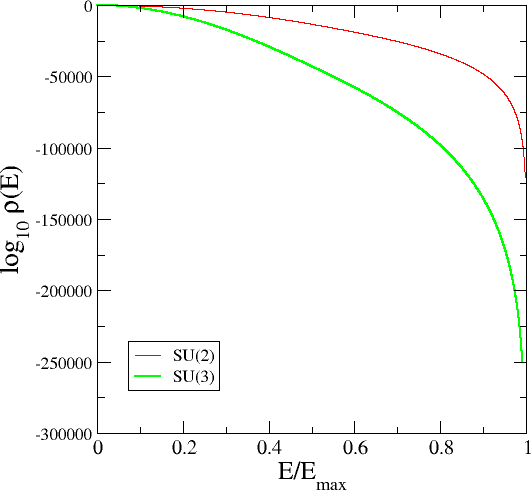}
\includegraphics[height=7cm]{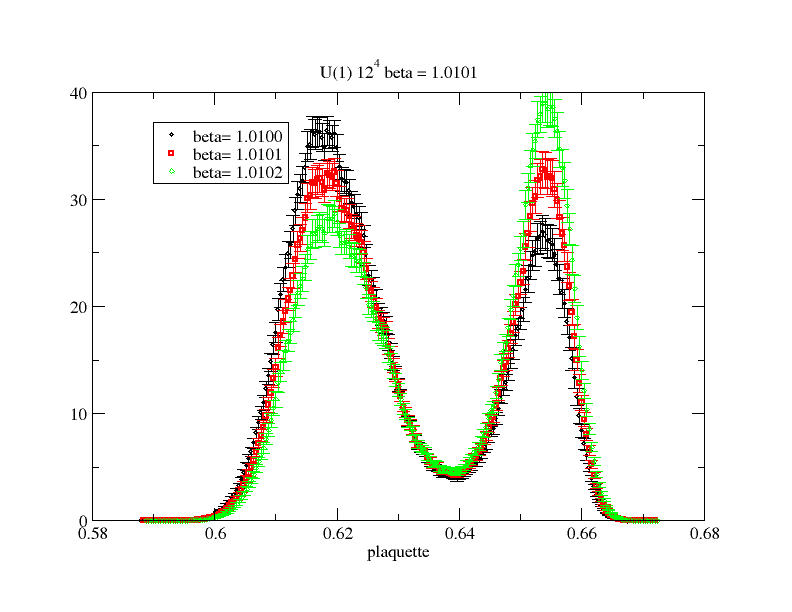}
\caption{ \label{fig:1} Left: The logarithm (base 10) of the density-of-states 
for a SU(2) and SU(3) lattice Yang-Mills theory for a $10^4$ lattice. Right: 
Probability distribution of the action for the compact U(1) theory for a 
$12^4$ lattice. 
}
\end{figure}
Figure~\ref{fig:1} shows an actual result for the density-of-states $\rho $
for the cases of a SU(2) and SU(3) gauge theory. It appears that $\log _{10}
\rho (E)$ is a remarkable smooth function over the whole action range 
$[0,E_\mathrm{max}]$ with $E_\mathrm{max}=60,000$ for the present case. 
This motivates a piecewise linear Ansatz for this quantity: 
\be 
\rho _{LLR} (E) = \rho (E_0) \, \exp \Bigl\{ 
a(E_0) \, (E-E_0) \Bigr\} , \; \; \; E_0 <E < E_0 + \delta E \; . 
\label{eq:15} 
\en
The task is now to find the expansion coefficients $a(E_0)$, which can be
interpreted as derivatives to the partition function: 
$$ 
a(E_0) \; = \; \frac{1}{Z} \; \frac{ dZ }{dE} \Big\vert _{E=E_0} \; . 
$$
Key ingredient to obtain $a(E_0)$ are truncated and re-weighted expectation
values~\cite{Langfeld:2012ah}: 
\bea 
{\langle \kern-.17em \langle} f\left(S[\phi]\right)  {\rangle \kern-.17em \rangle }(a)
&=& \frac{1}{\cal N} \int {\cal D}\phi \; f(S) \; \theta _{[E_0,\delta E]} \;
\; \mathrm{e}^{-a S[\phi] } \; ,
\label{eq:16} \\ 
\theta _{[E_0,\delta E]} &=& \left\{ \begin{array}{ll} 
1 & \hbox{ for } \; E_0 \le S[\phi] \le E_0 + \delta E \; , \\ 
0 & \hbox{else,}
\end{array}  \right. 
\hbo 
{\cal N} \; = \; \int {\cal D}\phi \;  \theta _{[E_0,\delta E]} \;
\; \mathrm{e}^{-a S[\phi] } \; .
\nonumber 
\ena
The double-bracket expectation values can be evaluated using standard
Monte-Carlo techniques: configurations are generated with respect to the 
Gibbs factor $\exp \{ - a S[\phi] \}$ and rejected if the target configuration 
would produce an action that falls outside the allowed action window. 
Using the definition of the density-of-states $\rho (E)$ in (\ref{eq:11}), we
can write: 
\be 
{\langle \kern-.17em \langle} f(S) {\rangle \kern-.17em \rangle }(a)
\; = \; \frac{1}{\cal N} \int {\rm d}E \; f(E) \; \rho (E) \, \theta _{[E_0,\delta E]} \;
\; \mathrm{e}^{-aE} \; . 
\label{eq:17}
\en
We now specialise to 
$$ 
f(S[\phi ]) \; = \; S[\phi ] \; - \; \left( E_0 + \frac{\delta E}{2} \right) 
=: \Delta S[\phi ] \; . 
$$
If the parameter $a$ in (\ref{eq:17}) equals the exact value $a_\mathrm{ex}$,
we obtain: 
\be 
{\langle \kern-.17em \langle} \Delta S[\phi ] {\rangle \kern-.17em \rangle
}(a_\mathrm{ex}) \; = \; \frac{1}{\cal N} \int {\rm d}E \; 
\left( E \; - \;  E_0 - \frac{\delta E}{2} \right) \; 
\theta _{[E_0,\delta E]} \; = \; 0 \; . 
\label{eq:18}
\en
The latter equation can be solved for $a_\mathrm{ex}$. If 
${\langle \kern-.17em \langle} \Delta S[\phi ] {\rangle \kern-.17em \rangle
}(a)$ is positive, the $a$ in the re-weighting factor $\mathrm{e}^{-a S[\phi]
}$  is too small to compensate the exact density-of-states $\rho $. 
This suggests the fixed-point iteration (with a suitable chosen relaxation 
parameter $\lambda >0$): 
\be 
a_{n+1} \; = \; a_n \; + \; \lambda \; {\langle \kern-.17em \langle} \Delta
S[\phi ] {\rangle \kern-.17em \rangle }(a_n) \; , \hbo 
a_\mathrm{ex} \; = \; \lim _{n \to \infty } a_n \; . 
\label{eq:19}
\en 
Practical details on the implementation of the LLR algorithms will be
presented in a forthcoming publication. The calculation of $a(E=E_0)$ is then
carried out for a range of action values $E_0$ and the the density-of-states 
$\rho (E)$ is recovered from (\ref{eq:15}). 
The probability distribution for the action $E$ is then easily estimated from 
$$ 
P(E; \beta ) \; = \; \rho (E) \, \exp\{ \beta E \} \; . 
$$
Error bars can be obtained by the standard bootstrap method. 
Given the set of $a$ values, this analysis takes little time on a standard
desktop making it feasible to explore a quasi-continuum of $\beta $ values. 
This is particularly interesting to locate a 1st order phase transition. 
Figure~\ref{fig:1}, right panel, shows our findings for the compact U(1) gauge
theory for a $12^4$ lattice. 

\medskip 
The above approach to the action density-of-states can be easily generalised 
to calculate the probability distribution $\rho (q)$ (\ref{eq:5}) for the
(integrated) Polyakov line $q$. To this aim, we set: 
$$ 
{\cal D} \phi \; = \; {\cal D} U_\mu \; \exp \{ S_\mathrm{YM}[U] \} \; .
$$
The LLR method now generates a quasi-continuum of probabilities 
$ \rho _\beta (q) \; \exp \{ j \, q \} $ as a function of the external source
  $j$. Note, however, that the generalised density-of-states $\rho _\beta (q)
  $ has to be re-calculated for each Wilson coefficient $\beta $.

\section{Numerical Results}
\begin{figure}
\includegraphics[height=6.6cm]{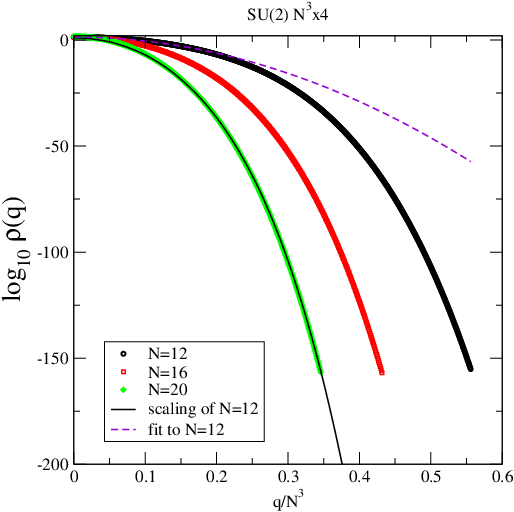} \hspace{0.5cm} 
\includegraphics[height=6.6cm]{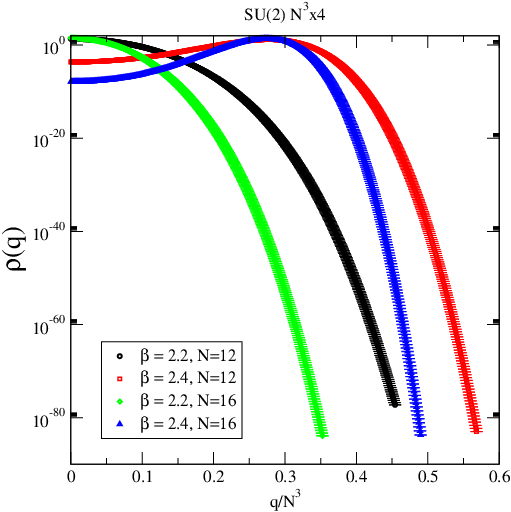}
\caption{ \label{fig:2} Left: The (integrated) Polyakov line probability 
distribution for SU(2) and a $N^3\times 4$ lattice. Right: 
Probability distribution for the confinement ($\beta =2.2$) phase and for 
the deconfinement phase ($\beta=2.4$). 
}
\end{figure}
\begin{figure}
\includegraphics[height=7cm]{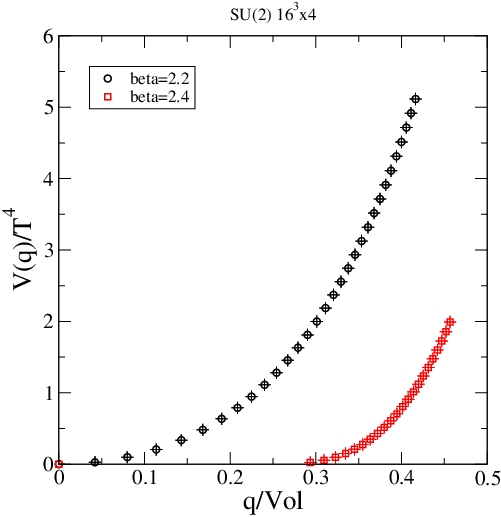}
\includegraphics[height=6.4cm]{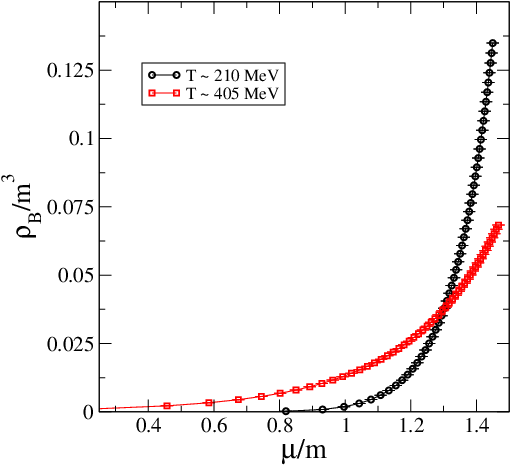}
\caption{ \label{fig:3} Left: The Coleman effective potential in units of the 
temperature $T$ for the SU(2) gauge theory for a $16^3 \times 4$ lattice 
for the confinement ($\beta =2.2$) and the high temperature ($\beta=2.4$) 
phase. Right: For SU(2), the ``baryon'' (diquark) density as a function of 
the chemical potential $\mu $ in units of the heavy quark mass chosen to 
match the charm quark in QCD. 
}
\end{figure}
The first simulations targeted the Polyakov line probability distributions 
$\rho _\beta (q)$ for a $N^3 \times 4 $ lattice for the confinement phase 
($\beta =2.2$). We have studied this distribution for several lattice 
sizes~\cite{Langfeld:2013xbf}. With the LLR algorithm, we quite easily 
obtain the distribution over hundreds of orders of magnitude (see figure 
\ref{fig:2}). Increasing the spatial value, decreases the width of the 
distribution. Also shown is a Gaussian fit to the data for $N=12$ (see dashed 
line in figure~\ref{fig:2}, left panel). We observe significant deviations 
from a Gaussian behaviour at large values $q$. We expect that 
$\log \rho(q)$ scales with the volume. To perform this consistency check using 
our numerical findings, we rescaled the data for $N=12$ to match with the data
for $N=20$. The re-scaled data (line in figure~\ref{fig:2}) nicely fall on top 
of the data for $N=20$ (green symbols). 

\medskip 
We then studied the probability distribution in the deconfinement phase 
by using $\beta =2.4$. The result is shown in figure~\ref{fig:2}, right panel. 
Due to spontaneous symmetry breaking, the distribution is suppressed for 
the ``false vacuum'' $q=0$. The suppression of $\rho (q=0)$ increases 
with increasing spatial volume resulting in a spontaneous breakdown of 
the centre symmetry in the infinite volume limit. 

\medskip 
Our numerical results benefit from the exponential error suppression of the 
LLR algorithm making it possible to calculate the Coleman effective potential 
(\ref{eq:3}) by a direct Legendre transformation, 
Figure~\ref{fig:3}, right panel, shows our findings for a $16^3\times 4$ 
lattice. In the confinement phase ($\beta=2.2$) the data points almost 
quadratically raise with increasing $q$ while in the high temperature 
phase ($\beta=2.4$) the potential is strongly suppressed for small $q$ 
(in fact, vanishing in the infinite volume limit due to spontaneous 
symmetry breaking). 

\medskip 
Finally, we show the ``baryon'', i.e., the diquark, density $\rho $ as 
a function of the chemical potential $\mu $ using the approximations leading 
to (\ref{eq:1},\ref{eq:2}). We have chosen ``physical'' parameters 
such as a string tension of $440 \, $MeV and a heavy quark mass matching that 
of the charm quark in QCD. In the confinement phase, we observe a sharp 
rise of the density when the quark chemical potential starts exceeding the 
quark mass gap. In the high temperature phase, we observe quite a significant 
density below the threshold. These contributions arise from temperature 
excitations of {\it single } quarks over the mass gap. This needs to be 
contrasted with the confinement phase where a diquark needs to be 
excited to obtain a contribution to the baryon density.

\end{document}